\documentclass[aps,prl,twocolumn,balance,superscriptaddress,floats]{revtex4}

\usepackage[a4paper, total={7.2in, 10in}]{geometry}
\usepackage{latexsym}
\usepackage{dcolumn}
\usepackage{amsmath}
\usepackage{epsf}
\usepackage{float}
\usepackage{hyperref}

\usepackage[pdftex]{graphicx}
\usepackage{epstopdf}
\usepackage{pdfpages}

\usepackage{upgreek}

\begin{document}


\title{On-chip Inter-modal Brillouin Scattering}

\author{Eric A. Kittlaus}
\affiliation{Department of Applied Physics, Yale University, New Haven, CT 06520 USA.}
\author{Nils T. Otterstrom}
\affiliation{Department of Applied Physics, Yale University, New Haven, CT 06520 USA.}
\author{Peter T. Rakich}
\affiliation{Department of Applied Physics, Yale University, New Haven, CT 06520 USA.}

\date{\today}

\begin{abstract}
Stimulated Brillouin interactions mediate nonlinear coupling between photons and acoustic phonons through an optomechanical three-wave interaction. Though these nonlinearities were previously very weak in silicon photonic systems, the recent emergence of new optomechanical waveguide structures have transformed Brillouin processes into one of the strongest and most tailorable on-chip nonlinear interactions. New technologies based on Brillouin couplings have formed a basis for amplification, filtering, and nonreciprocal signal processing techniques. In this paper, we demonstrate strong guided-wave Brillouin scattering between light fields guided in distinct spatial modes of a silicon waveguide for the first time. This inter-modal coupling creates dispersive symmetry breaking between Stokes and anti-Stokes processes, permitting single-sideband amplification and wave dynamics that permit near-unity power conversion. Combining these physics with integrated mode-multiplexers enables novel device topologies and eliminates the need for optical circulators or narrowband spectral filtering to separate pump and signal waves as in traditional Brillouin processes. We demonstrate 3.5 dB of optical gain, over 2.3 dB of net amplification, and 50\% single-sideband energy transfer between two optical modes in a pure silicon waveguide, expanding the design space for flexible on-chip light sources, amplifiers, nonreciprocal devices, and signal processing technologies.
\end{abstract}

\maketitle

\section{Introduction}

Stimulated Brillouin scattering (SBS) is a three-wave nonlinear process that produces coherent coupling between optical waves and acoustic phonons. Within waveguides, Brillouin interactions are remarkably tailorable, permitting a range of hybrid photonic-phononic signal processing operations that have no analogue in all-optical signal processing  \protect{\cite{Vidal07,Li2013,shinpper,marpaung,Kang2011,Huang11,kim2015,kang,braje09}}. Strong Brillouin interactions have only recently been created in silicon using a new class of optomechanical waveguides \protect{\cite{shinnatcomm,roel,van2015net,Kittlaus2016}} that produce record-high nonlinearities and admit intriguing dynamics \protect{\cite{kang,shinpper,marpaung,roel,Kittlaus2016}} as a basis for a variety of novel devices \protect{\cite{kang,shinpper,He15,marpaung}}. Further control of such Brillouin processes could enable phenomena such as mode cooling \protect{\cite{Bahl2012,1602.00205}} and nonreciprocal Brillouin scattering-induced transparency \protect{\cite{kim2015,Dong2015}}, and lead to new lasers, oscillators, filters \protect{\cite{Tanemura02,Zadok07,Vidal07,Wise11,zhang12}}, frequency sources \protect{\cite{Li2013,Merklein16}}, and signal processing technologies \protect{\cite{yao98,loayssa,olsson}} in silicon.

Many different types of Brillouin interactions are possible within microscale waveguides and devices, each producing a distinct set of phenomena. To date, strong forward stimulated Brillouin scattering (FSBS), also termed stimulated Raman-like scattering (SRLS) \protect{\cite{kang}}, has been achieved within silicon waveguides \protect{\cite{shinnatcomm,roel,van2015net,Kittlaus2016}}. Through FSBS, phonons mediate coupling between co-propagating light fields that are guided in the same optical mode. This interaction produces very strong optical nonlinearities ($10^4-10^5$ larger than in silica fibers)  \protect{\cite{shinnatcomm,roel}}, enabling large net amplification \protect{\cite{Kittlaus2016}} and cascaded energy transfer. This interaction also produces dynamics that are very different than those of the widely studied backward-SBS (BSBS) interaction \protect{\cite{agrawal2007nonlinear}, offering intriguing opportunities for new processes and phenomena. For example, FSBS is uniquely suited to hybrid photonic-phononic signal processing schemes based on phonon emit/receive operations that have recently been realized in silicon \protect{\cite{shinpper}. However, unlike BSBS, FSBS does not produce single-sideband gain. As a result, it is nontrivial in many cases to adapt BSBS device concepts and established techniques for signal processing \protect{\cite{yao98,loayssa,olsson}}, slow light\protect{\cite{okawachi05}}, and filtering \protect{\cite{Tanemura02,Zadok07,Vidal07,Wise11,zhang12} using FSBS interactions.

Alternatively, within multi-mode optomechanical waveguides it is also possible to create phonon-mediated coupling between light-fields guided in distinct spatial modes. This interaction, termed stimulated inter-modal Brillouin scattering (SIMS), creates an unusual symmetry-breaking between Stokes and anti-Stokes processes for forward-propagating waves. This form of symmetry breaking produces single-sideband gain while offering powerful new dynamics that enable a variety of unique operation schemes. SIMS and SIMS-like processes have recently been demonstrated in photonic crystal and nanoweb fibers \protect{\cite{kangprl,koehler16}}, admitting new physics compatible with both established technologies based on BSBS as well as many novel optical devices. Fiber experiments have used SIMS processes to create single-sideband amplification \protect{\cite{kangprl}}, self oscillation \protect{\cite{koehler16}}, and active optical isolation \protect{\cite{Kang2011}}, with dynamics that permit near-unity energy transfer between pump and signal waves \protect{\cite{kangprl}}. Inter-modal scattering in microsphere resonators permits attractive schemes for optical cooling \protect{\cite{Bahl2012}},  Brillouin scattering-induced transparency \protect{\cite{kim2015,Dong2015}} and nonreciprocal energy storage. All of these processes, and many others \protect{\cite{Tanemura02,Zadok07,Vidal07,yao98,braje09,loayssa,olsson,Wise11,Huang11,zhang12,Li2013}} become available on a silicon chip if engineerable forms of SIMS can be created.

In this paper, we demonstrate efficient and highly engineerable stimulated inter-modal Brillouin coupling on-chip for the first time. Through this process, a Brillouin interaction couples light fields that propagate in distinct spatial modes of a Brillouin-active silicon waveguide. This system decouples Stokes and anti-Stokes processes through symmetry breaking based on multimode dispersion. Harnessing this interaction, we demonstrate single-sideband optical amplification and unidirectional Brillouin energy transfer for the first time in silicon. Combining this novel form of Brillouin coupling with on-chip spatial mode-multiplexers offers a powerful new approach to Brillouin-based signal processing in silicon.

This interaction is realized in a novel integrated photonic/phononic waveguide system that permits net optical amplification and single-sideband energy transfer between two strong optical fields. Strong nonlinear coupling enables 3.5 dB of single-sideband small-signal gain, corresponding to 2.3 dB of net on-chip amplification in this low-propagation loss system. Pump- and signal-waves are coupled in and out of separate optical modes of a single Brillouin-active silicon waveguide using integrated mode multiplexers. This enables independent on-chip control of pump- and signal-waves without additional optical components such as isolators or narrowband filters. At higher guided-wave powers, this same device produces 50\% energy transfer between these two fields to produce stimulated mode conversion. Due to the inherent symmetry breaking of the inter-modal coupling, only two fields participate in this process, theoretically permitting near-unity power transfer between optical fields. This contrasts with FSBS-based energy transfer, where the amount of power transferred from one tone to another is fundamentally limited by stimulated comb-line generation. The realization of highly-engineerable SIMS in silicon unlocks a new platform for Brillouin interactions in silicon photonic systems. 

The unique physics of SIMS result from a form of coupling between phonons and light that differs from that of FSBS. Through FSBS, both Stokes and anti-Stokes scattering processes are mediated by the same phonon mode. Furthermore, the phase-matching condition for FSBS allows the same phonon to scatter light to many successive blue- and red-shifted orders \protect{\cite{kang,Kittlaus2016}}. By contrast, phase-matching for SIMS coupling to different optical modes requires that Stokes and anti-Stokes processes are mediated by distinct phonon modes. As a result, the dynamics of these processes become de-coupled, permitting single-sideband amplification as well as a variety of processes unique to inter-modal scattering \protect{\cite{kangprl,Kang2011}}.

\begin{figure*}[t]
\centering
\includegraphics[width=\linewidth]{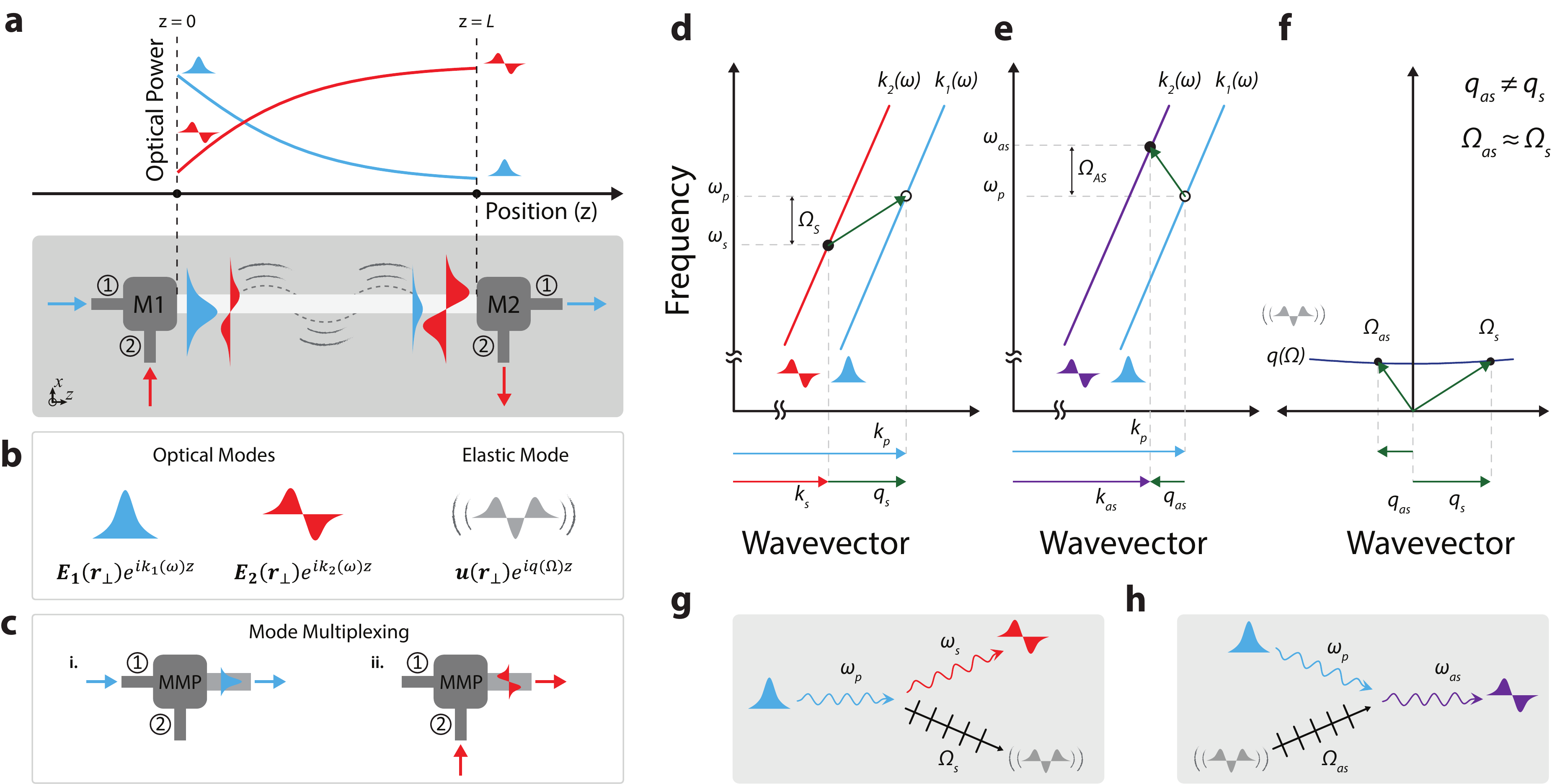}
\caption{On-chip inter-modal Brillouin scattering. (a) Operation scheme. Pump and Stokes waves are coupled into the fundamental and first excited modes of a waveguide through separate ports of an integrated mode multiplexer. While passing through the active device region, energy is transferred from pump to Stokes. At the end of the device, the two waves are demultiplexed through an identical mode multiplexer. (b) Schematic of the three modes participating in the Brillouin process. Two optical modes -- one even and the other odd in electric field symmetry -- are coupled through the interaction with an elastic mode with an even displacement profile. (c) Diagram of two-port mode multiplexer operation: Light injected into port 1 is coupled into the fundamental mode of a waveguide, whereas light coupled through port 2 is coupled into the same waveguide's first excited mode. (d-f) Diagrams showing dispersion relations for the participating modes. (d) depicts phase matching and energy conservation for a Stokes process, whereas (e) shown an anti-Stokes process. (f) plots the dispersion relation for the Brillouin-active acoustic mode. Note that phonons mediating Stokes and anti-Stokes processes have wavevector different in magnitude and sign, but nearly identical frequencies. (g) depicts a scattering diagram for a Stokes process, and (h) depicts a scattering diagram for an anti-Stokes process.}
\label{fig:xsbs}
\end{figure*}

\begin{figure*}[t]
\centering
\includegraphics[width=\linewidth]{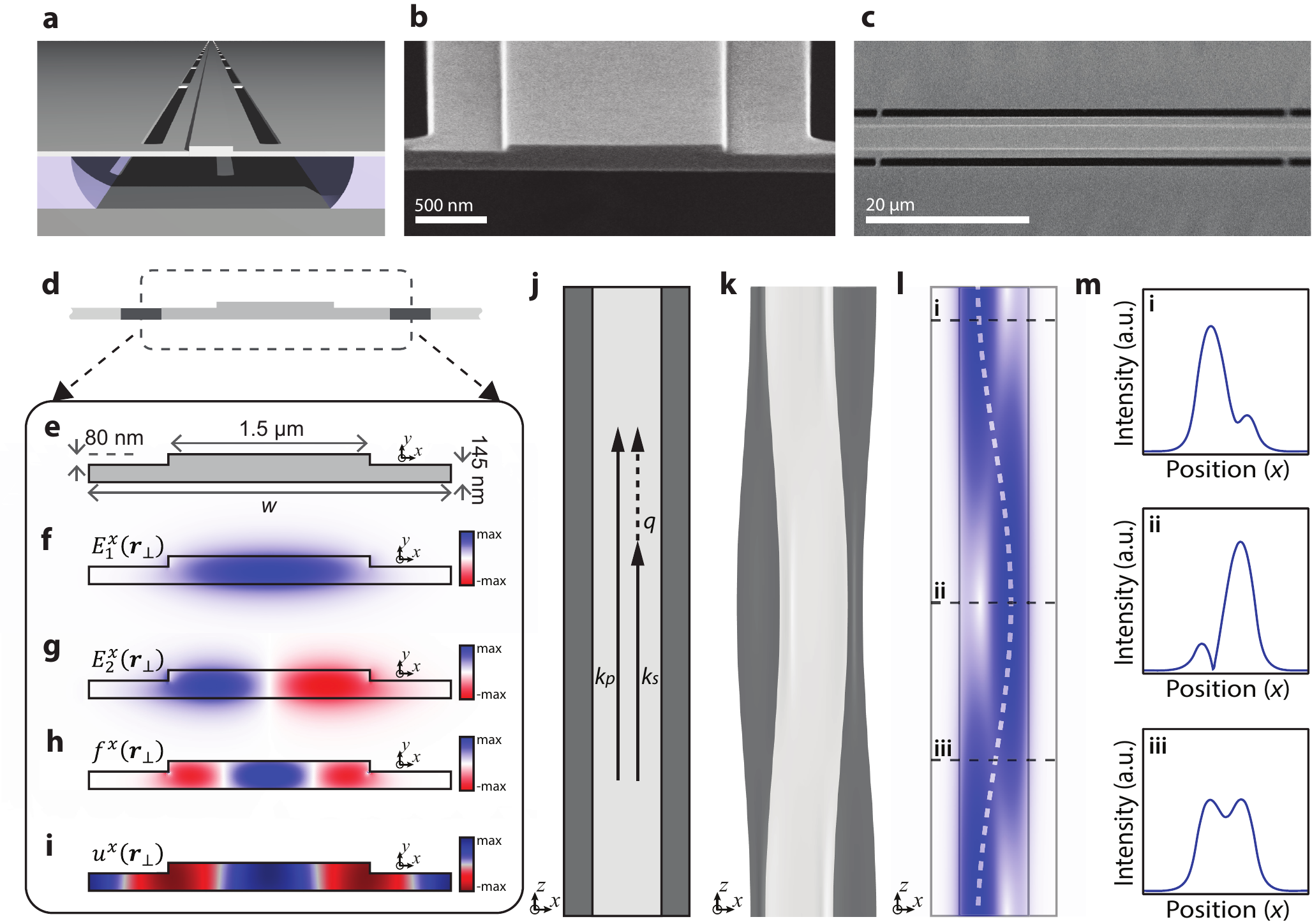}
\caption{SIMS-active waveguide. (a) Schematic of suspended Brillouin-active waveguide (b) SEM of device cross-section. The scale bar represents 500 nm in length. (c) Top-down SEM of suspended device, with a scale bar representing 20 $\upmu$m. (d) Diagram of device cross-section. Dashed region is plotted in (e) with relevant dimensions listed. (f) and (g) are $E_x$ fields of the first two guided optical modes. (h) $x$-component of the electrostrictive force generated by these optical modes. (i) $x$-displacement field of the $\sim$6 GHz Brillouin-active acoustic mode. (j), (k), and (l) sketch top-down views of a 14 $\upmu$m long section of the device including phase matching (j), elastic displacement for the membrane and ridge regions (k), and intensity beating of the two propagating optical modes (l). Three slices of the intensity profile in (l) are plotted in panels (m)i-iii.}
\label{fig:device}
\end{figure*}

\section{Operation Scheme}

We explore stimulated mode conversion and single-sideband amplification using the multi-port optomechanical system diagrammed in Fig. \ref{fig:xsbs}a. This system consists of a multi-mode hybrid photonic-phononic waveguide that is interfaced with two integrated mode multiplexers labeled M1 and M2.

We consider the interaction between two guided optical modes ${\bf E}_1(r_\perp)e^{ik_1(\omega)z}$ and ${\bf E}_2(r_\perp)e^{ik_2(\omega)z}$ and a Brillouin-active acoustic phonon mode ${\bf u}(r_\perp)e^{iq(\Omega)z}$, as sketched in \ref{fig:xsbs}b. Here ${\bf E}_j(r_\perp)$ and $k_j(\omega)$ are the electric field profile and wave-vector of the $j^{th}$ optical mode at frequency $\omega,$ and ${\bf u}(r_\perp)$ and $q(\Omega)$ are the elastic displacement field profile and wavevector of the acoustic mode at frequency $\Omega$.

Each port of the optical mode multiplexers, denoted as M1 and M2 in Fig. \ref{fig:xsbs}c, maps to a distinct spatial mode as shown in Fig. \ref{fig:xsbs}b. In the absence of nonlinear coupling, light injected into port 1 of M1 propagates in the symmetric mode $({\bf E}_1)$ of the optomechanical waveguide, exiting the system in port 1 of M2. Similarly, light entering port 2 of M1 propagates in the anti-symmetric mode $({\bf E}_2)$ of the waveguide, exiting the system in port 2 of M2. However, when pump- and signal-waves are injected into ports 1 and 2, respectively, resonant nonlinear coupling to the Brillouin-active phonon mode $({\bf u})$ scatters energy from the symmetric mode $({\bf E}_1)$ to the anti-symmetric mode $({\bf E}_2)$, producing both active mode conversion and single-sideband signal amplification.

This stimulated scattering process is driven by optical forces within the optomechanical waveguide segment. The pump wave $({\bf E}_1)$ interferes with the signal wave $({\bf E}_2)$ to produce a time-modulated optical force that excites a traveling-wave phonon $({\bf u}),$ which in turn scatters energy from pump to signal. This process, which is a type of stimulated inter-modal Brillouin scattering (SIMS), produces amplification and mode conversion through a stimulated Stokes process. In contrast to FSBS processes that have previously been demonstrated in silicon waveguides \protect{\cite{shinnatcomm,roel,van2015net,Kittlaus2016}}, this process also produces single-sideband amplification and unique forward scattering wave dynamics as a basis for new types of nonreciprocal devices \protect{\cite{Kang2011,kim2015,Dong2015}}.

Single-sideband amplification is possible in this system because the Stokes and anti-Stokes processes are no longer mediated by the same phonon mode. This is understood from the distinct phase matching conditions for these processes. Through the Stokes process, diagrammed in Fig. \ref{fig:xsbs}g, a pump photon ($\omega_p$, $k_1(\omega_p)$) guided in the symmetric mode $({\bf E}_1)$ scatters to a red-shifted photon ($\omega_s$, $k_2(\omega_s)$) guided in the anti-symmetric mode $({\bf E}_2)$, and a guided Stokes phonon ($\Omega_{s}$, $q_{s}$). For this process to occur, both energy conservation $(\Omega_s = \omega_p - \omega_s)$ and phase-matching $(q_{s} = k(\omega_p)-k(\omega_s))$ must be satisfied. Combined, these conditions require $q_{s} = k_1(\omega_p)-k_2(\omega_p-\Omega_{s})$. We use this condition as the basis for a succinct diagrammatic representation that includes both phase matching and energy conservation, as seen in Fig. \ref{fig:xsbs}d. A phonon that satisfies this condition must lie along the acoustic dispersion curve, $q(\Omega)$ sketched in Fig. \ref{fig:xsbs}f. This phase-matched phonon, which connects the initial (open circle) and final optical states (solid circle) sketched in Fig. \ref{fig:xsbs}d, is a \textit{forward} moving phonon.

Through the anti-Stokes process, diagrammed in Fig. \ref{fig:xsbs}h, the same pump photon ($\omega_p$, $k_1(\omega_p)$) combines with a guided phonon ($\Omega_{as}$, $q_{as}$) to produce a blue shifted photon ($\omega_{as}$, $k_2(\omega_{as})$). In this case, phase matching and energy conservation require $q_{as} = k_1(\omega_p+\Omega_{as})-k_2(\omega_p),$ which differs from the Stokes process. As seen from Fig. \ref{fig:xsbs}e, the anti-Stokes phonon that connects the initial optical state (open circle) and final optical state (solid circle) is a \textit{backward} moving phonon. 

Since the Stokes and anti-Stokes phonons are distinct (i.e., $q_s \neq q_{as}$) symmetry is broken between the two processes. As a result, the dynamics for the Stokes and anti-Stokes processes are de-coupled, permitting unidirectional single-sideband coupling between only two optical fields. (for further discussion see Supplementary Section S4.)

\section{Results}

This multi-port optomechanical system is fabricated from a single-crystal silicon layer using an SOI fabrication process (for details see Methods). The spatial mode multiplexers M1 and M2 address the individual modes of the optomechanical waveguide, and are fabricated on the same layer as the Brillouin-active waveguide segment. These mode multiplexers utilize asymmetric mode-selective directional couplers \protect{\cite{Dai13,Luo2014}} adapted to the low-loss ridge waveguide designs used here (see Supplementary Section S3 for details). Fiber arrays and grating couplers transfer light into single mode waveguides, which serve as the input- and output-ports of M1 and M2.

The Brillouin-active waveguide segment that supports stimulated inter-modal coupling is a suspended ridge waveguide of the type depicted in Fig. \ref{fig:device}a.  Scanning electron micro-graphs (SEMs) show the cross-section (Fig. \ref{fig:device}b) and top-view (Fig. \ref{fig:device}c) of the crystalline silicon device. The Brillouin-active waveguide is suspended over a 2.3 cm distance by an array of nanoscale tethers placed every 50 $\upmu$m (see Fig. \ref{fig:device}c). The active region consists of a 80 nm $\times$ 1.5 $\upmu$m wide ridge on a 135 nm thick silicon membrane of width $w = 2.85$ $\upmu$m. The cross-section of the active device region, highlighted in Fig. \ref{fig:device}d, is diagrammed in Fig. \ref{fig:device}e. 

Both light and sound are guided within the membrane-suspended waveguide of Fig. \ref{fig:device}e. Light is confined to the central ridge structure through total internal reflection, which guides the co-propagating TE-like optical modes. The symmetric $({\bf E}_1)$ and anti-symmetric $({\bf E}_2)$ spatial modes are plotted in Fig. \ref{fig:device}f-g at a wavelength of $\lambda=1550$ nm. Elastic waves are confined to this same structure due to the large acoustic impedance mismatch between silicon and air.

\begin{figure*}
\centering
\includegraphics[width=.807\linewidth]{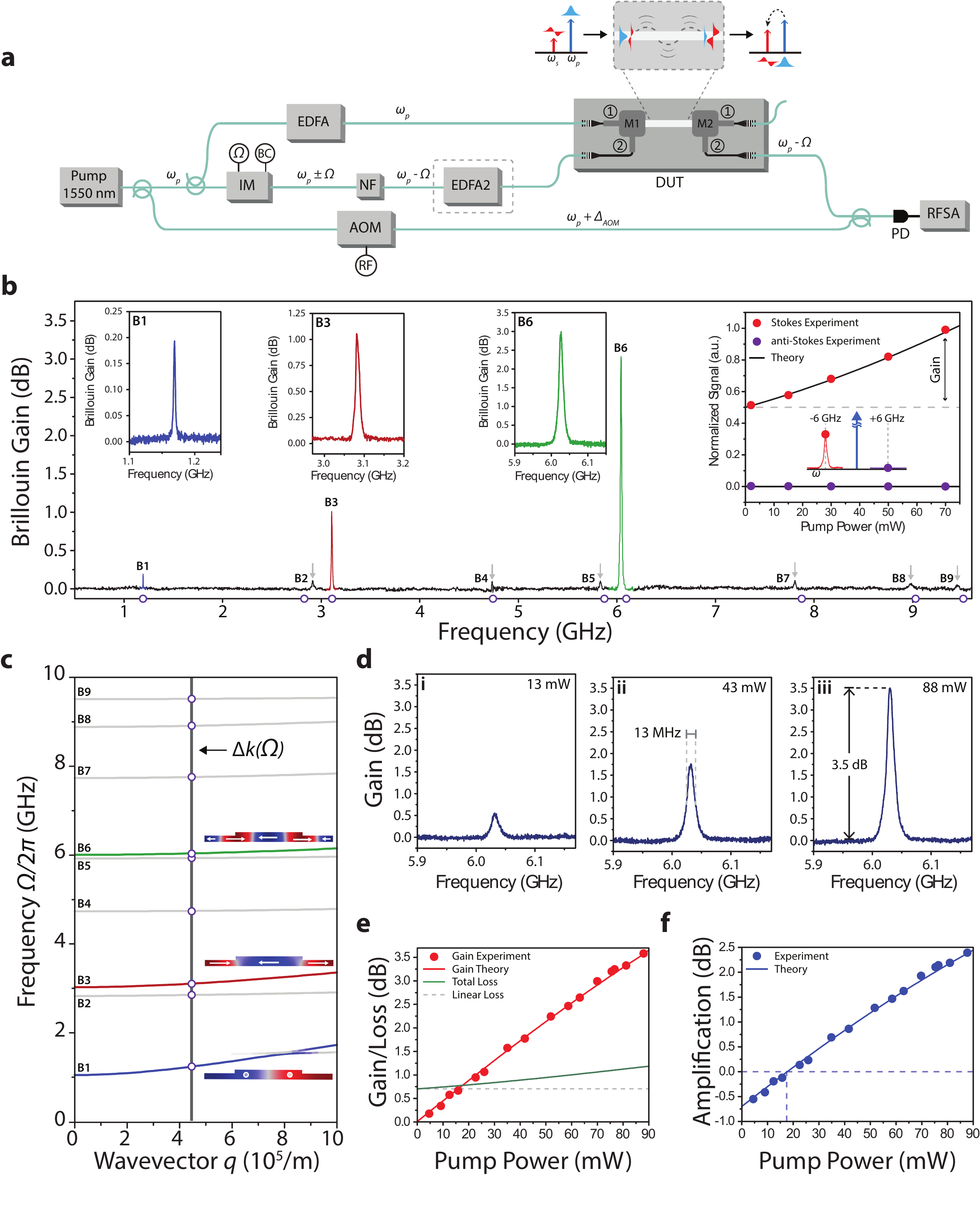}
\caption{Experimental results showing on-chip SIMS and net optical amplification. (a) Diagram of the experimental apparatus. A laser operating around 1550 nm is used to synthesize a strong pump wave through an EDFA while a separate branch is passed through an intensity modulator and notch filter to generate a frequency-shifted probe wave. An additional EDFA, denoted EDFA2, is inserted later for the optical energy transfer experiment. Pump- and probe-waves couple in and out of the device through mode multiplexers. After the device, the probe wave is combined with a frequency-shifted reference arm, and its intensity measured via heterodyne detection. (b) SIMS spectra over a 9-GHz span showing several Brillouin-active acoustic modes, with three highlighted and plotted in more detail in insets. Simulated frequencies from multi-physics simulations are denoted by violet circles along the abscissa. A fourth inset shows single-sideband gain as a function of pump power when driving acoustic resonance B6--note that while Stokes light experiences gain, no light is scattered to the anti-Stokes order. A data trace at high pump power is included schematically, showing single-sideband gain as the probe wave is swept through the Brillouin resonance. Red and purple dots correspond to the measured gain when driven on-resonance. (c) Calculated dispersion curves for the observed Brillouin-active acoustic modes. Displacement is diagrammed for modes B1, B3, and B6 next to their respective curves. The vertical grey line plots the wavevector of Brillouin active phonons under the conditions tested. (d)-(f) show Brillouin gain and amplification data tested for the strongest Brillouin-active mode B6. Panels i, ii, and iii of (d) plot SIMS Brillouin gain spectra obtained for three different pump powers. These data show a narrowband Brillouin resonance at 6.03 GHz. (e) plots peak gain (red), linear loss (dash) and total loss (green) versus on-chip pump power. (f) Net on-chip amplification from (e), calculated by subtracting total loss from optical gain. These data are measured with the pump wave in the fundamental optical mode and the probe wave in the higher-order mode.}
\label{fig:ssb}
\end{figure*}

This system supports numerous guided elastic waves with longitudinal, shear, and flexural character. Of these, only a small number are Brillouin-active and mediate transfer of energy between modes $({\bf E}_1)$ and $({\bf E}_2)$. In addition to phase matching, Brillouin coupling requires appreciable overlap between optical force density,  produced by the interference between $({\bf E}_1)$ and $({\bf E}_2)$, and the elastic displacement field. One such Brillouin active phonon mode that produces strong inter-modal Brillouin coupling at 6 GHz frequencies is shown in Figs.  \ref{fig:device}i and  \ref{fig:device}l. Figure \ref{fig:device}i shows the dominant (x-component) of elastic displacement field (${\bf u} ({r}_\perp)$) within the waveguide cross-section, and Figure \ref{fig:device}l is a top-view showing elastic distortion of this same mode. Comparing the electrostrictive optical force distribution of Fig. \ref{fig:device}h, with Fig. \ref{fig:device}i, we see that the transverse displacement produces good overlap with the optical forces to mediate coupling.

\subsection{Stimulated Inter-Modal Scattering}

We examine the inter-modal Brillouin response of this system using nonlinear laser spectroscopy. The stimulated inter-modal Brillouin scattering spectrum is obtained by measuring phonon-mediated energy transfer between the symmetric and anti-symmetric optical modes of the Brillouin active waveguide segment. A strong pump-wave ($\omega_p$) and weak signal-wave ($\omega_s = \omega_p-\Omega$) enter the symmetric (${\bf E}_1$) and anti-symmetric (${\bf E}_2$) waveguide modes through ports 1 and 2 of mode multiplexer M1, respectively; power guided in these same modes exit ports 1 and 2 of multiplexer M2. As the pump-probe detuning ($\Omega$) is swept through resonance, Brillouin coupling produces energy transfer between pump- and probe-waves propagating in the symmetric and anti-symmetric modes, yielding the resonant features seen in the gain spectrum Fig. \ref{fig:ssb}b; throughout this article, we refer to the relative change in the probe intensity as the Brillouin gain.

These Brillouin gain measurements are conducted using the apparatus of Fig \ref{fig:ssb}a. Both the pump- and probe-waves are synthesized from the same continuous-wave laser. The laser output ($\omega_p$) is amplified using an erbium doped fiber amplifier (EDFA) to form the pump-wave. A single frequency probe-wave ($\omega_p-\Omega$) is synthesized from the same laser output using an electrically driven intensity modulator and an optical notch filter as seen in Fig.\ref{fig:ssb}a. The frequency separation, $\Omega$, between the pump- and probe-waves is controlled with sub-Hertz precision using a microwave signal generator. 
Since the probe wave is kept to $< 200$ $\upmu$W, pump depletion is negligible. 

At the device output, the optical waves exiting port 2 of M2 of the chip are analyzed using wideband heterodyne spectral analysis. As shown in Fig \ref{fig:ssb}a, signal-light exiting port 2 is combined with an optical local oscillator ($\omega_p+\Delta_{AOM}$)  generated by frequency shifting the laser output. As the signal and local oscillator interfere, the intensity of each tone within the signal-wave is observed as a unique beat-note using a fast photodiode and a spectrum analyzer. The Brillouin gain spectrum of Fig. \ref{fig:ssb}b is obtained by measuring the power contained in the output probe-wave at frequency $\omega_p-\Omega$, as the pump-probe detuning $\Omega$ is varied from 500 MHz to 9.5 GHz (for more details see Supplementary Section S2).

The frequencies of the identified Brillouin-active phonon modes, labeled B1-9 in Fig. \ref{fig:ssb}b, show good agreement with frequencies predicted through multi-physics simulations (denoted by violet circles along the abscissa). The phononic dispersion curves (computed based on measured device dimensions) for each of the Brillouin-active phonon modes are highlighted in Fig \ref{fig:ssb}c. Phase-matched coupling occurs where the phonon wave-vector ($q_m(\Omega)$) of the $m^{th}$ Brillouin-active  dispersion curve matches the optical wave-vector mismatch, $\Delta k(\Omega) =k_1(\omega_p)-k_2(\omega_p -\Omega) $; these frequencies are identified by the intersection between the phonon dispersion curves and the vertical line (grey) (for further details on the acoustic modes of this system, see Supplementary Section S7).

The gain spectrum (Fig. \ref{fig:ssb}b) reveals dominant Brillouin resonances, labeled B1, B3, and B6, corresponding to Brillouin-active phonon modes with frequencies of 1.18, 3.09, and 6.03 GHz. Insets within Fig. \ref{fig:ssb}c illustrate the dominant displacement character of each mode. The simulated coupling strengths, obtained through full-vectorial finite element simulations of the type described in Ref. \protect{\cite{wenjun}}, produce good agreement with the observed Brillouin nonlinearity of each mode; these simulations reveal that the interaction is primarily mediated by photoelasticity.

\subsection{Dispersive Symmetry Breaking}

Next we examine stimulated inter-modal Brillouin dynamics through both gain and nonlinear power transfer measurements; throughout these studies, we study these dynamics using the 6.03 GHz resonance that exhibits the largest Brillouin gain--this resonance is labeled B6 in Fig. \ref{fig:ssb}b.
A key characteristic of SIMS through coupling to these phonon modes is single-sideband gain. To investigate these dynamics, we perform power dependent spectral analysis of the probe signal exiting port 2 of M2 as we vary the pump wave power. Figure \ref{fig:ssb}d, shows the spectral content of the transmitted waves obtained through heterodyne measurements when pump and probe light is coupled through the Brillouin-active mode. When the probe detuning is set to the Brillouin resonance frequency and the pump power is varied, we observe single-sideband gain, plotted in the rightmost inset of Fig. \ref{fig:ssb}b. Here the measured Stokes (red dots) and anti-Stokes (purple dots) powers are shown as a function of pump power. In contrast to FSBS processes, as the pump power is increased from 0 to 70 mW and the Stokes (red-shifted) sideband experiences gain, no light is scattered to the anti-Stokes order (for further discussion see Supplementary Section S4).

These measurements establish that this inter-modal coupling produces the predicted dispersive symmetry breaking necessary to enable single-sideband gain and support existing schemes for slow light \protect{\cite{okawachi05}} and Brillouin-based optical memory \protect{\cite{Zhu07}}.  Next, we show that this system produces net optical amplification, necessary to support new laser geometries and robust integrated photonic performance; a similar figure of merit is required for efficient Brillouin-scattering induced transparency \protect{\cite{kim2015}}.

\subsection{Single-Sideband Amplification}

Net on-chip optical amplification requires that the Brillouin gain exceeds both linear and nonlinear propagation losses. To quantify total inter-modal gain and net optical amplification, we perform power-dependent measurements of the gain and loss experienced by the probe wave. Using the experiment diagrammed in Fig. \ref{fig:ssb}a, the pump-probe detuning $\Omega$ is swept through the Brillouin frequency $\Omega_S$ to measure the resonant Brillouin gain spectrum. Three such SIMS gain spectra, for incident on-chip fundamental-mode pump-wave powers of 13, 43, and 88 mW, respectively, are shown in Fig. \ref{fig:ssb}d.i-iii. These data show a high-quality factor ($Q = \Omega_S/\Delta\Omega = 460$) Brillouin resonance at $\Omega_S = 6.03$ GHz and 3.5 dB of Brillouin gain at the highest tested pump powers. Peak gain vs. pump power, as well as measured linear and nonlinear loss, are plotted in Fig. \ref{fig:ssb}e. Net optical amplification, calculated by subtracting total loss from Brillouin gain, is plotted in Fig. \ref{fig:ssb}f. 2.3 dB of amplification is achieved at the highest tested pump power of 88 mW. These data are fit to a complete nonlinear Brillouin gain model to obtain a Brillouin gain coefficient of $G_B = 470 \pm  30$ W$^{-1}$m$^{-1}$. The measured gain and frequency agree well with simulated values of $G_B = 410 \pm  70$ W$^{-1}$m$^{-1}$ and $\Omega_S = 6.07$ GHz for this $w = 2.85$ $\upmu$m device.

The SIMS process can also be pumped in the higher-order mode and amplify light in the fundamental mode--data for four different pump-probe configurations are plotted in Supplementary Section S6. 

Significant Brillouin-based optical amplification in silicon relies on low optical propagation losses and large Brillouin gain. In this waveguide system, linear losses were determined through length-dependent ring-resonator finesse measurements and nonlinear losses were measured through power-dependent transmission measurements (for full details see Supplementary Section S2). Due to reduced spatial overlap between the waveguide modes, nonlinear losses impacting SIMS are 40-70\% lower than those for intra-modal scattering, resulting in lower power-dependent losses than those in the FSBS system of Ref. \protect{\cite{Kittlaus2016}. This is important as it permits low-threshold amplification and robust operation at high pump powers. Greatly reduced fifth-order losses due to free carrier effects, in particular, are necessary to support high-power laser designs.

\subsection{Inter-Modal Energy Transfer}

Strong inter-modal Brillouin coupling permits significant nonlinear power transfer between optical fields and may enable new on-chip optical devices. These dynamics have previously been observed only in highly-nonlinear photonic crystal fiber, where light is scattered between distinct polarization states \protect{\cite{kangprl}}. This interaction was used as the basis for new forms of active optical isolators \protect{\cite{Kang2011}}. The operation scheme of this active isolator system was made possible through the use of polarization multiplexing which behaves analogously to on-chip mode multiplexing.

In principle, SIMS processes of the type realized here can enable similar forms of nonreciprocal optical devices including circulators and switches on a silicon chip. To explore the possibility of such operations in an integrated photonic platform, we quantify nonlinear energy transfer in the large-signal regime (i.e. where significant energy transfer occurs).

We quantify inter-modal Brillouin energy transfer in the large-signal regime using the apparatus of Fig \ref{fig:ssb}a; here, an additional EDFA (labeled EDFA2) is introduced to boost the probe wave intensity. Pump- and probe-waves of equal intensity are injected into modes ${\bf E}_1$ and ${\bf E}_2$ of the Brillouin-active waveguide, respectively, with pump-probe frequency detuning ($\Omega$) equal to the Brillouin resonance frequency of 6.03 GHz. As the pump and probe waves propagate within the active device region, the pump wave $({\bf E}_1)$ is scattered into mode ${\bf E}_2$ causing pump depletion and probe-wave gain. As before, pump and probe waves enter and exit the active device region through optical mode multiplexers M1 and M2.

Fig. \ref{fig:et}b plots energy transfer data as a function of combined incident power, with a maximum of 75 mW in both pump and probe waves, for a device with width $w = 2.77$ $\upmu$m and nonlinear coefficient $G_B = 420 \pm  20$ W$^{-1}$m$^{-1}.$ At the highest tested powers, 50\% energy transfer from pump to Stokes is demonstrated. In principle, energy transfer efficiencies approaching 100\% are possible with either a longer Brillouin-active region or higher pump power \protect{\cite{kangprl,Chen14}}. 

These results demonstrate a dramatically different regime of operation from the small signal-gain limit. In contrast to the undepleted pump regime, a significant fraction (many tens of milliwatts) of power is transferred between the participating optical fields. This result also highlights the high power-handling of the unclad silicon membrane waveguide--up to 150 mW powers are guided without adverse effects. The specific dynamics of this two-wave nonlinear energy transfer process contrast with those of FSBS, where arbitrarily many optical fields participate in the energy transfer process (see Supplementary Section S5). Finally, the good agreement achieved between experiment and theory corroborate measurements of nonlinear gain and loss.

The highly-controllable dynamics of SIMS-mediated energy transfer may be applied to create new nonlinear physics on a silicon chip. In addition to new schemes for nonreciprocal devices \protect{\cite{Kang2011}}, this process supports significant total power transfer even over a short propagation length. Longer propagation lengths ($\sim$ 7 cm) with similar gain and loss coefficients or different operation schemes could permit $>90\%$ energy transfer \protect{\cite{Chen14}}.  

\begin{figure}[t]
\centering
\includegraphics[width=\linewidth]{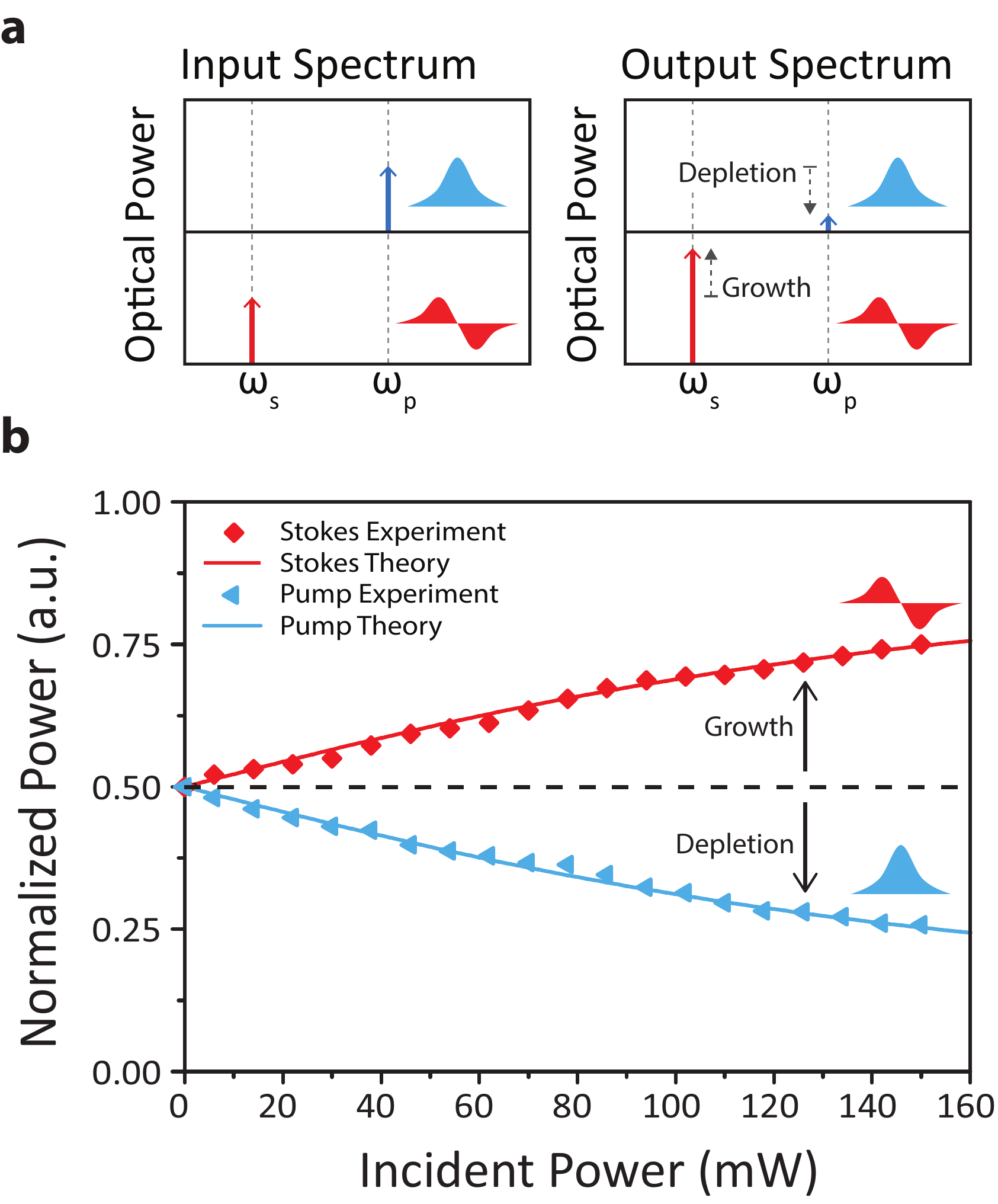}
\caption{Inter-modal Brillouin energy transfer. (a) Diagrams of input and output spectra for the energy transfer experiment. The total change in normalized signal after Brillouin coupling is denoted graphically by dashed arrows in the output spectrum. (b) Energy transfer fraction as a function of total incident power for one Brillouin-active device.}
\label{fig:et}
\end{figure}

\section{Discussion}

We have shown that this new type of on-chip Brillouin scattering produces dispersive symmetry breaking, net optical amplification, and appreciable mode conversion. These characteristics permit a variety of processes not previously possible in silicon photonics.

Dispersive symmetry-breaking between Stokes and anti-Stokes processes through Brillouin interactions in silicon permits numerous new operations. As discussed above, due to symmetry-breaking SIMS supports single-sideband optical amplification and energy transfer. Since Stokes and anti-Stokes scattering are mediated by different phonons, several schemes for mode cooling are now possible. In addition to resonator-based cooling schemes \protect{\cite{Bahl2012}}, the high nonlinear coupling of this system compares favorably with predictions of necessary gain-power products to observe spontaneous Brillouin cooling in a linear waveguide \protect{\cite{1602.00205}}. The large Brillouin gain and low propagation loss of the membrane waveguide system may also permit the creation of Brillouin scattering-induced transparency on a silicon chip \protect{\cite{kim2015,Dong2015}}. Finally, symmetry-breaking enables the adaptation of many traditional technologies based on BSBS, such as existing methods for slow light \protect{\cite{okawachi05}}, beam combining \protect{\cite{Rodgers99}}, and optical memory \protect{\cite{Zhu07}} which are not readily achieved through on-chip FSBS. 

The combination of these highly-tailorable physics with broadband mode multiplexers \protect{\cite{Dai13,Luo2014}} offers many intriguing possibilities for integrated photonic systems. In contrast with previous approaches to Brillouin scattering, the operation scheme used for SIMS in silicon eliminates the need for circulators or narrowband filters to separate pump and signal waves--the four-port system discussed here allows automatic multiplexing/demultiplexing of these waves into spatially separate waveguides which can then be routed to other devices on the same chip. The four-port linear-waveguide system can also be adapted into a laser design by connecting ports 2 of M1 and M2 as diagrammed in Fig. \ref{fig:xsbs}a. This design creates a cavity for Stokes light while being transparent for pump light. In contrast with traditional Brillouin lasers, this removes the requirement that the FSR match the Brillouin resonance frequency, allowing Brillouin lasers of arbitrary footprint on a silicon chip. The gain and power handling demonstrated in the SIMS-active waveguide should be adequate to create such a laser.

Beyond this specific system, SIMS also enables new devices based on cross-modal coupling on a silicon chip. SIMS is a form of active mode conversion, which may have applications in active switching, power routing, or on-chip mode-division multiplexing \protect{\cite{Dai13,Luo2014}}. The physics of inter-modal Brillouin scattering can be generalized to any number of spatial modes with different symmetries and couplings, in contrast with stimulated inter-polarization scattering. The SIMS-active membrane waveguide also allows geometric tuning of the gain spectrum--the resonant frequency of the Brillouin-active phonons is directly related to the phononic membrane width, unlike in most BSBS systems \protect{\cite{shinnatcomm}}. Furthermore, the wavevector and velocity of Brillouin-active phonons can be tuned through optical mode engineering, allowing highly-selectable phonon excitation for on-chip acousto-optic device applications.

In conclusion, we have demonstrated stimulated inter-modal Brillouin scattering in an on-chip system for the first time. This new nonlinear coupling allows inter-modal amplification, single-sideband energy transfer, and unprecedented control over the Brillouin interaction. Through independent photonic and phononic control, we have demonstrated ultralow nonlinear losses intrinsic to the inter-modal coupling and robust acoustic performance. Using this system, we demonstrated net amplification of 2.3 dB  and 50\% energy transfer from one optical field to another. This tailorable Brillouin nonlinearity can support a wide range of hybrid photonic-phononic technologies for RF and photonic signal processing, and is readily integrable in silicon photonic systems. This work extends the growing body of research on integrated Brillouin photonics by adding powerful new control over the spatial behavior and nonlinear dynamics of the Brillouin interaction.

\section{Methods}

\subsection{Device Fabrication}
The silicon waveguides were written on a silicon-on-insulator chip with a 3 $\upmu$m oxide layer using electron beam lithography on hydrogen silsesquioxane photoresist. Following development, a Cl$_2$ reactive ion etch (RIE) was employed to etch the ridge waveguide structure. After a solvent cleaning step, slots were written to expose the oxide layer, again with electron beam lithography of ZEP520A photoresist and Cl$_2$ RIE. The device was then wet released via a 49\% hydrofluoric acid etch of the oxide undercladding. The waveguide structures under test are each comprised of 461 suspended segments.

\subsection{Experiment}
Both experiments used a pump laser operating around 1550 nm. The following abbreviations are used in the experimental diagram: 

Fig. \ref{fig:ssb}a: IM Mach-Zehnder intensity modulator, BC DC bias controller, EDFA erbium-doped fiber amplifier, NF notch filter, DUT device under test, AOM acousto-optic frequency shifter, PD photodetector, RFSA radio frequency spectrum analyzer.  

For further details on the experimental setup, see Supplementary Section S2.

\subsubsection{Acknowledgements}
This work was supported through a seedling grant under the direction of Dr. Daniel Green at DARPA MTO and by the Packard Fellowship for Science and Engineering; N.T.O. acknowledges support from the National Science Foundation Graduate Research Fellowships Program. We thank Prashanta Kharel for technical discussions involving Brillouin dynamics, and Dr. Michael Rooks and Dr. Michael Power for assistance with process development. We also thank Shai Gertler for his assistance in the development of early mode multiplexer designs.

\subsubsection{Author Contributions}
E.A.K. and N.T.O. fabricated the waveguide devices. E.A.K., N.T.O., and P.T.R. developed multi-physics simulations for and designed the devices. E.A.K. and N.T.O. conducted experiments with the assistance of P.T.R. E.A.K., P.T.R., and N.T.O. developed analytical models to interpret measured data. All authors contributed to the writing of this paper.

\subsubsection{Additional Information}
\noindent \textbf{Competing financial interests:} The authors declare no competing financial interests.

\bibliographystyle{naturemag-ed} 
\bibliography{cites}
\newpage

\clearpage


\section*{Supplementary material}

# On-Chip Inter-modal Brillouin Scattering

Eric A. Kittlaus[1], Nils T. Otterstrom[1], and Peter T. Rakich[1]

[1]*Department of Applied Physics, Yale University, New Haven, CT 06520 USA.*

## S1. Nonlinear Coefficients and Loss Model

We calculate the effective guided wave nonlinear coefficients using an electromagnetic-wave finite-element model (FEM) simulation. These nonlinear coefficients are enhanced compared to their bulk values due to the nontrivial vectorial nature of the highly-confined optical modes [S1]. Using the method of Ref. [S1] and a bulk value of $n_2 = 4.5 \times 10^{-18}$ m$^2$W$^{-1}$ [S2], we calculate guided wave Kerr coefficients of $\gamma_k^{11} = 74 \pm 11$ m$^{-1}$W$^{-1}$, $\gamma_k^{22} = 71 \pm 11$ m$^{-1}$W$^{-1}$, and $\gamma_k^{12} = \gamma_k^{21} = 44 \pm 7$ m$^{-1}$W$^{-1}$, where the superscripts *ij* correspond to the intra- and inter- modal nonlinear coefficients for the fundamental and first higher order guided modes, respectively. Using a bulk value for the two-photon coefficient of $\beta_{\text{TPA}} = 7.9 \times 10^{-12}$ mW$^{-1}$ we find $\beta^{11} = 34 \pm 10$ m$^{-1}$W$^{-1}$, $\beta^{22} = 30 \pm 9$ m$^{-1}$W$^{-1}$, and $\beta^{12} = \beta^{21} = 20 \pm 6$ m$^{-1}$W$^{-1}$. Here we use units of guided-wave power, i.e. $\gamma_k \approx n_2 \omega / (c \cdot A_{eff})$ and $\beta \approx \beta_{\text{TPA}} / A_{eff}$, where $A_{eff}$ is the effective nonlinear mode area.

We characterize the nonlinear loss characteristics of our system through measurements of intra- and inter-modal nonlinear loss for each mode fit to a nonlinear propagation loss model. The intra-modal propagation loss in a single-mode silicon waveguide system is generally captured by the differential equation

$$\frac{dP_i}{dz} = -\alpha^i P_i - \beta^{ii} P_i{}^2 - \gamma^{iii} P_i{}^3 .$$

Here $\alpha^i$ is the linear loss coefficient for mode $i$ and $\gamma^{iii}$ is the intra-modal nonlinear loss coefficient due to TPA-induced free carrier absorption (FCA). Fitting measured nonlinear transmission measurements (Fig. S1a,c) to this model allows the measurement of $\gamma^{iii}$.

In the presence of a bright pump $P_i$, the inter-modal nonlinear loss experienced by a weak field $P_j$ in a different mode is given by the additional equation

$$\frac{dP_j}{dz} = -\alpha^j P_j - 2\beta^{ji} P_i P_j - \gamma^{jii} P_i^2 P_j .$$

Comparing this equation to measured cross-modal nonlinear loss with the calculated value for $\beta^{ji}$ allows determination of $\gamma^{jii}$.

Linear loss $\alpha^i$ is determined by measuring the intrinsic Q-factor of ring resonators fabricated from multimode silicon waveguides of different lengths. The results of these studies (Plotted in Figure S1b,d) determine linear propagation losses to be $0.24 \pm 0.02$ dB/cm ($\alpha^1 = 5.5 \pm .5$ m$^{-1}$) for the fundamental mode and $0.30 \pm 0.12$ dB/cm ($\alpha^2 = 6.8 \pm 2.7$ m$^{-1}$) for the higher-order mode. To measure free carrier losses, power-dependent transmission data were fit to the nonlinear loss models using Mathematica and assuming the values of $\beta^{11}$, $\beta^{12}$, and $\beta^{22}$ calculated in Section S1. The intra-modal data plotted in Figure S1c-d are best fit with $\gamma^{111} = 900 \pm 400$ m$^{-1}$W$^{-2}$ and $\gamma^{222} = 720 \pm 430$ m$^{-1}$W$^{-2}$. These data are consistent with a free carrier lifetime of 1.2 ns according to a Drude-Sommerfeld model of free carrier absorption [S3] or a lifetime of 1.5 ns according to a Drude-Lorentz model [S4,5]. According to the overlap integral method of Ref. [S3], $\gamma^{211}$ is expected to be 41% of $\gamma^{111}$ in magnitude. Measurements of small signal cross-modal nonlinear loss of .45 dB at 76 mW pump powers are consistent with $\gamma^{211} \approx \gamma^{122} = 310 \pm 200$ m$^{-1}$W$^{-2}$, which is around 34% of $\gamma^{111}$, consistent with this calculation.

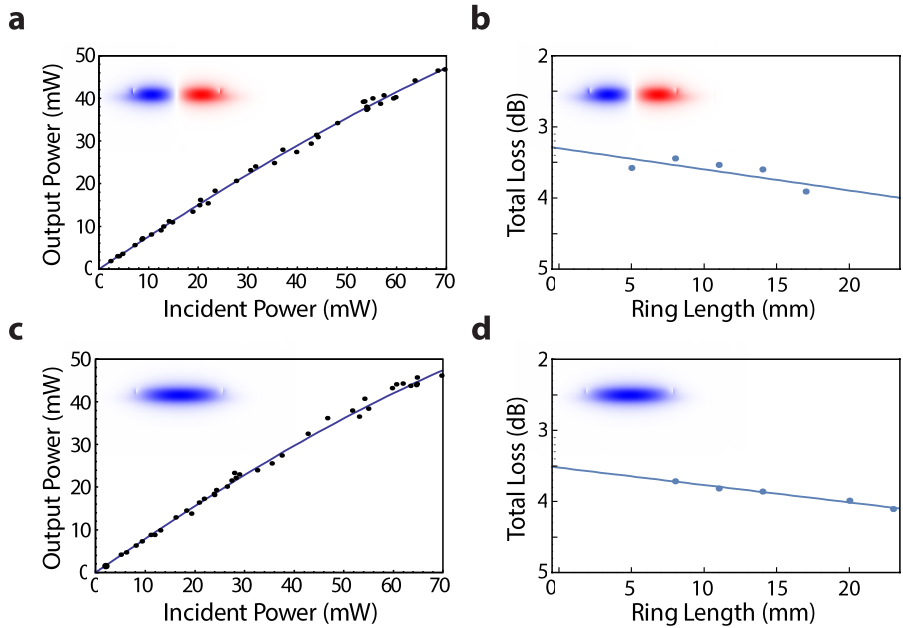

**Figure S1.** Linear and nonlinear loss measurements at $\lambda \approx 1550$ nm for the membrane waveguide. Panels (a) and (b) plot input vs. output power for the device under study and loss vs. ring resonator length for the anti-symmetric mode $E_1$. (c) and (d) plot these data for the fundamental mode $E_0$.

## S2. Brillouin Gain Model and Measurement

In the Brillouin amplification experiment, a strong pump field is used to amplify a weak Stokes signal. In the small signal limit ($P_p \gg P_s$), the coupled differential equations relating the propagating powers in the two fields in the presence of nonlinear loss are[S3]

$$\frac{1}{P_p(z)}\frac{dP_p(z)}{dz} = -\left(\alpha^p + \beta^{pp}P_p(z) + \gamma^{ppp}P_p^2(z)\right)$$

$$\frac{1}{P_s(z)}\frac{dP_s(z)}{dz} = -\alpha^s + \left(G_B(\Omega) - 2\beta^{sp} - \gamma^{spp}P_p(z)\right)P_p(z).$$

Here $\alpha^i$, $\beta^{ij}$ and $\gamma^{ijk}$ are the loss coefficients discussed in the previous section with the superscripts $p$ and $s$ referring to the indices of the pump and Stokes modes. $P_p(z)$ and $P_s(z)$ are the pump and Stokes powers at position $z$ along the device length, and $G_B(\Omega)$ is the inter-modal Brillouin gain coefficient. The device acts as a linear amplifier for the Stokes signal, with total amplification dependent on the input pump power.

After passing through the active device and mode demultiplexer, probe light is coupled with light from the original pump laser frequency-shifted by an acoustic-optic modulator to frequency $\omega_{LO} = \omega_p + \Delta_{AOM}$. (For these experiments, $\Delta_{AOM} = 44$ MHz.) While the power $P_{LO}$ in this field is kept constant throughout all measurements, the Stokes power has a frequency dependency as the pump-probe detuning is swept through the Brillouin resonance.

$$P_S = P_S^0 + \Delta P_S(\Omega)$$

The power in the Stokes field is measured via the power of the RF beat-note at frequency $\Omega + \Delta_{AOM}$ between Stokes and reference fields ($\propto P_s P_{LO}$) incident on a fast photodetector.

As the Stokes probe is swept through the Brillouin resonance, the resulting RF power measured is:

$$P_{RF}(\Omega + \Delta_{AOM}) = \eta P_{LO}\left(P_S^0 + \Delta P_S(\Omega)\right).$$

Here $\eta$ is a coefficient of proportionality related to the detector efficiency. To determine the change in Stokes power, this signal is normalized as

$$\frac{P_{RF}(\Omega + \Delta_{AOM})}{\eta P_{LO}P_S^0} = 1 + \frac{\Delta P_S(\Omega)}{P_S^0}.$$

The measured data from the Brillouin gain experiment are fit to this model to determine the Brillouin gain coefficient $G_B = 470 \pm 30$ m$^{-1}$W$^{-1}$. The gain coefficient is numerically simulated to be $G_B = 410 \pm 70$ m$^{-1}$W$^{-1}$ through a full-vectorial simulation with the commercially available finite-element method software COMSOL according to the method Ref. [S6], with values for the photoelastic tensor components for silicon ($p_{11}, p_{12}, p_{44}$) = (–0.09, 0.017, –0.051) [S7]. As shown in Fig. 2 of the main article, the Brillouin coupling is mediated primarily by the *x*-component of the electrostrictive force, with negligible contributions due to radiation pressure.

### S3. Mode Converters

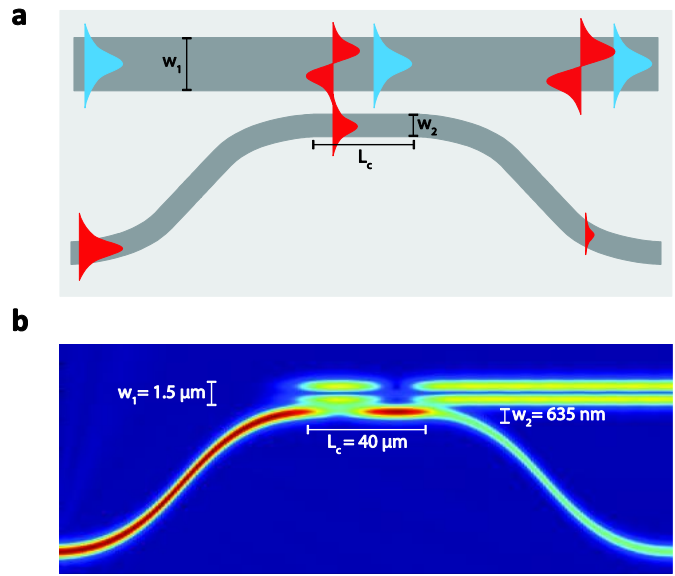

**Figure S2.** (a) Schematic of the mode selective directional coupler. (b) Simulation of coupling into the first excited mode with typical dimensions. In this particular simulation, 78% of the light is converted into the first excited spatial mode.

We employ mode-selective directional couplers to multiplex light into two distinct spatial modes of the hybrid photonic-phononic waveguide [S8]. Before the mode multiplexer, pump and Stokes light are spatially separated, each guided by a single-mode 450 nm wide rib waveguide. The waveguide guiding the pump field is tapered to a width of 1.5 μm which supports multimode operation. Simultaneously, the waveguide guiding the Stokes light is tapered to a width (635 nm) such that its effective index closely matches that of the first excited mode of the 1.5 μm waveguide. As depicted in Fig. S2, the narrow waveguide then channels the Stokes wave towards the multimode waveguide to permit evanescent coupling into the higher order mode.

From a coupled mode theory formulation [S9], one can show that the fraction of light coupled ($C$) into the higher order mode is

$$C = \frac{\kappa_{12}}{\beta_0} \sin(\beta_0),$$

$$\beta_0 = \sqrt{\kappa_{21}\kappa_{12} + \frac{(\beta_1 - \beta_2)^2}{2}}.$$

Here $\kappa_{12}$ and $\kappa_{21}$ are the coupling coefficients and $\beta_1$ and $\beta_2$ are the propagations constants of the modes of interest. The above equation shows that near unity power transfer is possible given identical effective indices. It also highlights that both the magnitude and phase of the coupling change dramatically with effective index mismatch, permitting strong coupling into the desired mode with little crosstalk to other spatial modes.

Using the process outlined in the methods section, we fabricate the mode multiplexers on a silicon on insulator platform. These directional couplers achieve 60 to 95 percent coupling from the fundamental mode of the 635 nm waveguide to the first excited mode of the multimode waveguide depending on fabrication parameters, with a typical value of around 75%.

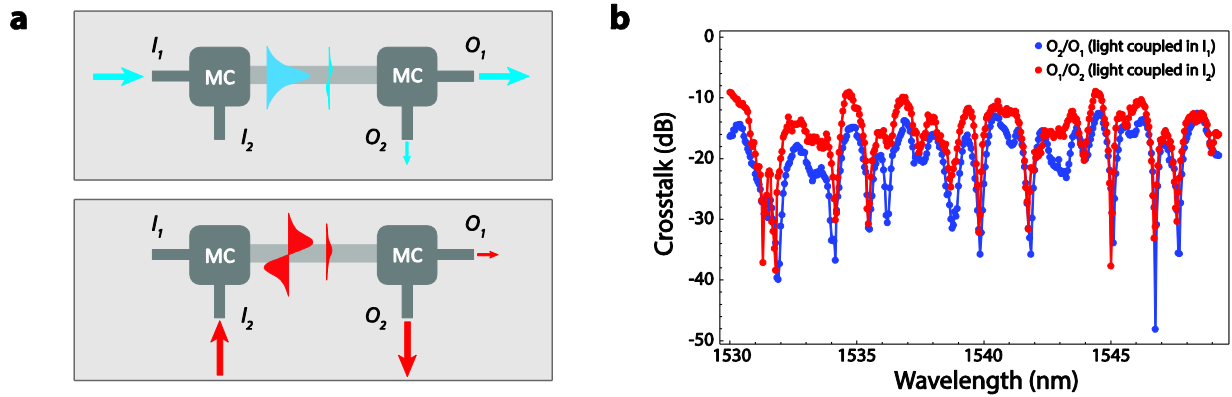

**Figure S3.** (a) Schematic for crosstalk measurement: Light injected into input port (I-1) is coupled into the fundamental mode with a small amount of light in the first excited mode. Light coupled into input port (I-2) is converted into the first excited mode with some residual light in the fundamental. (b) Typical crosstalk of a multiplexer and demultiplexer in series. Crosstalk is calculated by dividing powers of the light from the two output ports (O-1 and O-2).

After passing through the active device region, an identical coupler demultiplexes the two modes into two spatially separated single-mode waveguides. Any residual light (light not demultiplexed) in the first excited mode of the main waveguide is removed by a spatial mode

filter at the output. Ideally, each port of the multiplexer couples uniquely to its respective mode. In practice, some small amount of light is (de)multiplexed into the incorrect mode or port. Figure S3 shows a schematic for the crosstalk measurement and the crosstalk for a typical mode multiplexer and demultiplexer in series. Median crosstalk values for this configuration are typically -15 to -20 dB, though all experiments are carried out in regions with total crosstalk less than -20 dB.

## S4. SIMS vs. FSBS Dynamics

In the case of forward intra-modal SBS, the same phonon is phase-matched to both Stokes and anti-Stokes processes. As a result, the dynamics of these fields are coupled, and the equations of motion for the participating optical and acoustic envelope fields in a waveguide can be written as follows [S10]:

$$\frac{\partial \bar{B}}{\partial t} + v_0 \frac{\partial \bar{B}}{\partial z} = i(\Omega - \Omega_B)\bar{B} - i\big(g_0^* \bar{A}_s^* \bar{A}_p + g_1^* \bar{A}_p^* \bar{A}_{as}\big),$$

$$\frac{\partial \bar{A}_p}{\partial t} + v_p \frac{\partial \bar{A}_p}{\partial z} = -i\big(g_o \bar{A}_s \bar{B} + g_1 \bar{B}^* \bar{A}_{as}\big),$$

$$\frac{\partial \bar{A}_s}{\partial t} + v_s \frac{\partial \bar{A}_s}{\partial z} = -i g_0^* \bar{B}^* \bar{A}_p,$$

$$\frac{\partial \bar{A}_{as}}{\partial t} + v_{as} \frac{\partial \bar{A}_{as}}{\partial z} = -i g_1 \bar{A}_p \bar{B}.$$

Here $\bar{B}$, $\bar{A}_p$, $\bar{A}_s$, and $\bar{A}_{as}$ are the envelope functions for the phonon field and pump, Stokes, and anti-Stokes fields, respectively. $v_0$, $v_p$, $v_s$, and $v_{as}$ are the corresponding group velocities, $\Omega = \omega_p - \omega_s$ is the pump/Stokes detuning and $\Omega_B$ is the Brillouin resonance frequency, and $g_o$ and $g_1$ are the Brillouin coupling strengths for Stokes and anti-Stokes processes, respectively, and are assumed to be equal here. Since Stokes and anti-Stokes processes couple to the pump through the same phonon field, appreciable light is scattered from the pump to both fields.

By contrast, through SIMS Stokes and anti-Stokes processes couple through phonons with distinct wavevector. As a result, the two processes are de-coupled, and the equations of motions for the envelope functions can be written as follows [S10]:

$$\frac{\partial \bar{B}_s}{\partial t} + v_0 \frac{\partial \bar{B}_s}{\partial z} = i(\Omega - \Omega_s)\bar{B}_s - ig_0^* \bar{A}_s^* \bar{A}_p,$$

$$\frac{\partial \bar{A}_p}{\partial t} + v_p \frac{\partial \bar{A}_p}{\partial z} = -ig_o \bar{A}_s \bar{B}_s,$$

$$\frac{\partial \bar{A}_s}{\partial t} + v_s \frac{\partial \bar{A}_s}{\partial z} = -ig_0^* \bar{B}_s^* \bar{A}_p,$$

with a similar set of equations for anti-Stokes scattering:

$$\frac{\partial \bar{B}_{as}}{\partial t} + v_0 \frac{\partial \bar{B}_{as}}{\partial z} = i(\Omega - \Omega_{as})\bar{B}_{as} - ig_1^* \bar{A}_p^* \bar{A}_{as},$$

$$\frac{\partial \bar{A}_p}{\partial t} + v_p \frac{\partial \bar{A}_p}{\partial z} = -ig_1^* \bar{B}^* \bar{A}_{as},$$

$$\frac{\partial \bar{A}_{as}}{\partial t} + v_{as} \frac{\partial \bar{A}_{as}}{\partial z} = -ig_1 \bar{A}_p \bar{B}.$$

Note that the phonon mediating the anti-Stokes process, $\bar{B}_{as}$, is different than the phonon $\bar{B}_s$ mediating Stokes coupling. These equations produce the single-sideband optical coupling inherent to SIMS.

## S5. Energy Transfer model

In the case where the pump and Stokes waves are similar in magnitude, additional nonlinear terms must be added to the SIMS power evolution equations [S3]:

$$\frac{1}{P_p(z)}\frac{dP_p(z)}{dz} = -\left(\alpha^p + \beta^{pp}P_p(z) + \gamma^{ppp}P_p^2(z)\right) - (2\beta^{ps} - G_B(\Omega) + 4\gamma^{pps}P_p(z) + \gamma^{pss}P_s(z))P_s(z)$$

$$\frac{1}{P_s(z)}\frac{dP_s(z)}{dz} = -\left(\alpha^s + \beta^{ss}P_s(z) + \gamma^{sss}P_s^2(z)\right) - (2\beta^{sp} + G_B(\Omega) + 4\gamma^{ssp}P_s(z) + \gamma^{spp}P_p(z))P_p(z)$$

Multiple strong waves present in a nonlinear waveguide experience additional loss due to cross-field losses. This effect hinders energy transfer efficiency at high powers, suggesting that longer devices operating at lower powers or larger effective mode areas are amenable to highly-efficient nonlinear energy transfer.

It is worth noting that the coupling for this form of nonlinear energy transfer involves only two participating fields. This contrasts with intra-modal FSBS, where strong coupling results in stimulated energy transfer to multiple successive orders. Data for different input powers is plotted for the present work (SIMS) and for intra-modal FSBS in Fig S4.a-b.

Energy transfer through traditional FSBS energy is intrinsically limited to approximately 50% due to energy transfer to higher order comb lines [S11,S12]. This effect results from the FSBS-active phonon phase-matching to arbitrarily many Brillouin processes. The phonon that mediates SIMS phase matches unidirectionally from one optical mode to another, precluding any cascading effects and permitting in principle complete energy transfer from one field to another.

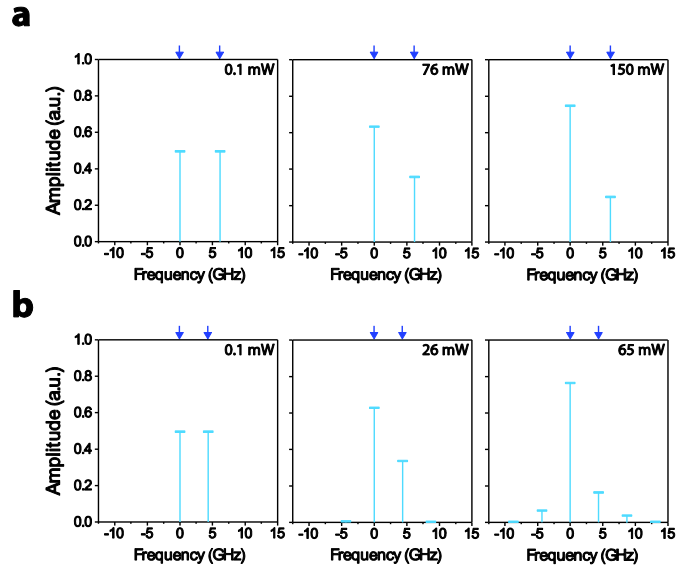

**Figure S4.** Experimental data at different total incident powers showing nonlinear energy transfer for SIMS (a) and FSBS (b). The FSBS data are from Ref. [S12]. The blue arrows above each panel denote where two equal-intensity drive tones are injected into the waveguide. While SIMS only couples light between two optical fields, at high powers FSBS produces cascaded energy transfer.

## S6. Stokes and anti-Stokes Data

The gain data in the main text are for a signal wave guided in the first-excited optical mode of a ridge waveguide while pump light is guided in the fundamental mode. The opposite configuration is also possible and supports comparable amplification. Conversely, a small signal

that is blueshifted from a strong pump by the Brillouin resonance frequency experiences attenuation through an anti-Stokes process—though in fact this is no different than the strong pump depletion limit, or the near-unity power transfer limit of the energy transfer experiment from the main article. Figure S5 plots data for these four pump/probe configurations in a Brillouin active waveguide with $w = 2.77$ μm, $G_B = 404 \pm 20$ m$^{-1}$W$^{-1}$, and $\Omega_B = 6.16$ GHz. This experiment used asymmetric Y-junction spatial mode converters which allow narrowband near-unity coupling into each optical mode [S13].

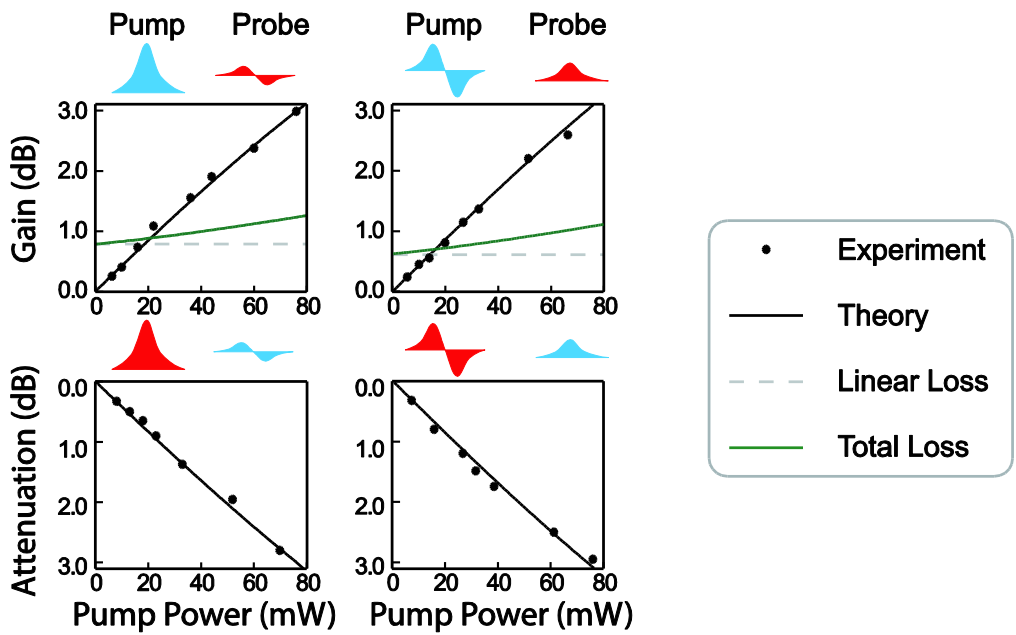

**Figure S5.** Plots of Stokes gain and anti-Stokes attenuation for a probe signal in four-different pump-probe configurations. The top left panel corresponds to the pump/probe configuration presented in the main article. The top two panels show gain for a weak probe red-shifted from the pump by the Brillouin resonance frequency, while the bottom two panels show attenuation of a weak probe which is blue-shifted from the pump by the Brillouin resonance frequency.

## S7. Elastic Mode Simulations

Fig. 2 in the main text depicts acoustic dispersion for the Brillouin-active modes observed. There are many other resonant modes of the membrane waveguide system which do not couple

appreciably through SIMS. Complete acoustic dispersion relations for frequencies $\Omega/2\pi < 10$ GHz and wavevector $q < 10^6$ m$^{-1}$ is plotted in Fig S6.

The Lamb wave-like elastic modes of the system can be separated into four categories based on their direction of motion. The first type is modes experiencing compression and rarefaction in the $x$-direction (red lines in Fig. S6). The 3 and 6 GHz modes which couple most strongly to the optical waveguide modes through SIMS are of this type. The displacement field of an example mode at around 4.3 GHz is plotted in Fig. S6.i.

The second family of modes are asymmetric Lamb wave-like modes with in-plane motion in the $x$- and $y$-directions, as shown by the 3.5 GHz mode plotted in Fig. S6.ii. These elastic modes are similar to transverse acoustic waves on a string or thin membrane. The third family of modes experiences compressive displacement in the longitudinal ($z$) direction. An example mode at 4.7 GHz is plotted in in Fig. S6.iii. Finally, three modes with linear dispersion correspond to the fundamental transverse $x$- and $y$-displacement modes and the $z$-compressive mode.

These modes are simulated using a full-vectorial elastic wave method using the commercially-available finite element model package COMSOL.

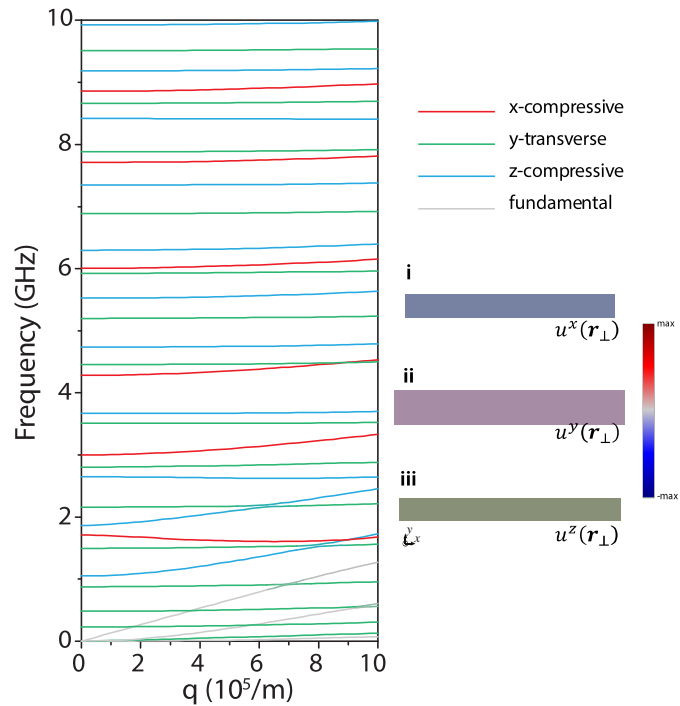

**Figure S6.** Acoustic dispersion relations for GHz-frequency guided acoustic modes of the suspended membrane waveguide. The modes are categorized according to their displacement field and examples of three types are plotted in panels i-iii.

## S8. Brillouin Resonance Frequency Tuning vs. Device Width

The frequencies of the guided acoustic waves in the Brillouin-active membrane waveguide are directly related to device dimension. By lithographically varying the dimensions of the system, this allows frequency tuning of the Brillouin gain spectrum. Figure S7 depicts theory and data for the frequencies of the two strongest Brillouin-active resonances around 3 and 6 GHz for a range of device widths. The predicted frequencies are calculated using COMSOL as in the previous section. Good agreement is achieved between experiment and theory over the tested range of dimensions.

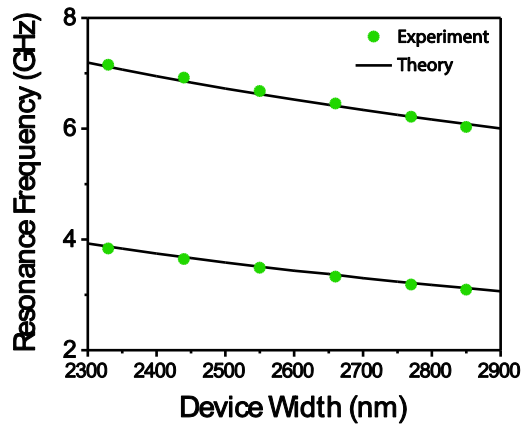

**Figure S7.** Simulated and measured acoustic frequencies for the two acoustic modes with the largest Brillouin gain as a function of device width. Good agreement is achieved between experiment and theory.

## S9. Sensitivity to Dimensional Broadening

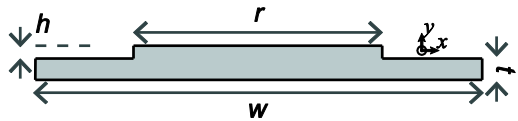

**Figure S8.** Device cross-section labeling the various dimensions which specify the device design.

As has been previously studied, the suspended ridge waveguide design supports a relatively high tolerance to dimensionally-induced broadening of the Brillouin resonance lineshape. In contrast to traditional forward-SBS, SIMS is significantly more sensitive to variations in device thickness and ridge waveguide height due to an additional dispersion-related broadening

mechanism: As studied in Ref. [S14], there are two dominant sources of inhomogeneous broadening of the gain spectrum in nanoscale Brillouin-active waveguides. The first is direct broadening via acoustic frequency shifting of the resonance as device dimensions vary along the propagation length. The second results from variations of the optical wavenumber that cause frequency shifts via the phase matching condition. Traditional FSBS, which couples to a phonon with vanishingly small wavevector, is devoid of the second broadening mechanism. Since the SIMS process studied here has a nonvanishing wavevector, this additional issue could lead to a degradation in measured effective quality factor. However, the acoustic quality factors for the SIMS phonon mode compare well with existing FSBS systems, boding well for the potential of further system scaling.

Simulated sensitivities in resonance frequency of the 6 GHz acoustic mode for small changes in device dimension (labeled in Fig. S8) are tabulated below. These calculated values take into account both potential broadening mechanisms.

| Dimension | Description | Sensitivity to Dimension Error |
|---|---|---|
| *w* | Total membrane width | $1.7 \times 10^6$ Hz/nm |
| *t* | Membrane Thickness | $1.7 \times 10^6$ Hz/nm |
| *r* | Ridge waveguide width | $9.2 \times 10^5$ Hz/nm |
| *h* | Ridge waveguide height | $1.4 \times 10^6$ Hz/nm |
| | Ridge offset from center | $<<10^5$ Hz/nm |



\end{document}